\documentclass[12pt,a4paper,psamsfonts]{article}
\usepackage{amsmath,amssymb,amsthm}
\usepackage{eucal}
\usepackage[hidelinks]{hyperref}

\numberwithin{equation}{section}

\usepackage{geometry}
\def\a{\alpha}

\def\d{\delta}
\def\e{\epsilon}

\def\l{\lambda}

\def\s{\sigma}

\def\G{\Gamma}

\def\Om{\Omega}

\newcommand{\hph}[1]{{\hphantom{#1}}}
\def\o{\over}

\def\lp{\left(}
\def\rp{\right)}
\def\ls{\left[}
\def\rs{\right]}

\newcommand{\SU}{\operatorname{SU}}
\newcommand{\be}{\begin{equation}}
\newcommand{\ee}{\end{equation}}
\newcommand{\non}{\nonumber}
\newcommand{\Tr}{\operatorname{Tr}}
\newcommand{\p}{\partial}
\newcommand{\al}{\alpha}
\newcommand{\bea}{\begin{eqnarray}}
\newcommand{\eea}{\end{eqnarray}}
\newcommand{\w}{\wedge}
\newcommand{\la}{\lambda}

\makeatletter
\renewcommand\section{\@startsection {section}{1}{\z@}%
                                   {-3.5ex \@plus -1ex \@minus -.2ex}
                                   {2.3ex \@plus.2ex}%
                                   {\normalfont\large\bfseries}}
\renewcommand\subsection{\@startsection{subsection}{2}{\z@}%
                                     {-3.25ex\@plus -1ex \@minus -.2ex}%
                                     {1.5ex \@plus .2ex}%
                                     {\normalfont\bfseries}}
\makeatother

\begin{document}

\begin{center}
\addtolength{\baselineskip}{.5mm}
\thispagestyle{empty}
\begin{flushright}
{\sc MI-TH-1534}\\
\end{flushright}

\vspace{20mm}

{\Large  \bf The $\alpha'$-Expansion of Calabi-Yau Compactifications}\footnote{Dedicated to A.~Strominger on the occasion of his 60th birthday. To appear in ICCM notices.}
\\[15mm]
{Katrin Becker, Melanie Becker and Daniel Robbins}
\\[5mm]
{\it George P. and Cynthia W. Mitchell Institute for }\\
{\it Fundamental Physics and Astronomy, Texas A\& M University,}\\
{\it College Station, TX 77843-4242, USA}\\[5mm]

\vspace{20mm}

{\bf  Abstract}
\end{center}
We consider $\alpha'$-corrections to Calabi-Yau compactifications of type II string theory.  These were discussed from the string worldsheet approach many years ago in terms of supersymmetric non-linear $\s$-models by Nemeschansky and Sen as well as Gross and Witten.
There it was shown that once $\a'$-corrections are included, the internal manifold solving the string equations of motion is still Calabi-Yau though not Ricci flat.  In this brief note we review these results and compare with a space-time effective field theory approach, in which we show that $\SU(3)$-holonomy manifolds become $\SU(3)$-structure manifolds once such corrections are included.

\vfill
\newpage


\section{Introduction}

Ever since the discovery of Calabi-Yau compactifications of string theory~\cite{Candelas:1985en}, physicists have wondered, if corrections to these manifolds would spoil the property, that K\"ahler manifolds with vanishing first Chern class solve the string equations of motion
\be
dJ=0,\qquad c_1(M_6)=0.
\ee
Here $M_6$ denotes the six-dimensional internal manifold, $J$ is the K\"ahler form and $c_1(M_6)$ is the first Chern class of $M_6$.  The vanishing of the first Chern class is a topological condition that holds, in particular, if $M_6$ admits a Ricci-flat metric
\be
R_{ab}=0,\qquad\mathrm{with}\qquad a,b=1,\cdots,6.
\ee

Many years ago Calabi conjectured~\cite{MR0085583}, that a compact K\"ahler manifold with vanishing first Chern class  admits a K\"ahler metric with $\SU(3)$ holonomy in the same K\"ahler class. This conjecture was subsequently proven in the famous work by Yau~\cite{Yau:1977ms}.

By definition, it is always possible to find a covariantly constant spinor $\eta$ on $\SU(3)$ holonomy manifolds
\be
\nabla_a\eta=0.
\ee
This implies that $M_6$ is Ricci-flat as can be shown in two simple steps~\cite{Candelas:1985en}. First, the commutator gives
\be
\ls\nabla_a,\nabla_b\rs\eta=0\quad\Longrightarrow\quad R_{abcd}\G^{cd}\eta=0.
\ee
Here $R_{abcd}$ is the 6D Riemann tensor and $\G_a$ are the 6D gamma matrices.  Second, contracting with $\G^b$ shows Ricci flatness of $M_6$
\be
R_{ab}=0.
\ee
How do string theory corrections influence this result?  This is the subject of this note.

\subsection{Wordsheet versus Space-time Approach}

There are two approaches that can be followed to describe the perturbatively corrected manifold:
\begin{enumerate}
\item Worldsheet approach in terms of an $\mathcal{N}=2$ supersymmetric nonlinear $\s$-model. This was used by Nemeschansky and Sen and Gross and Witten many years ago~\cite{Nemeschansky:1986yx,Gross:1986iv} and is described in the next section.
\item The space-time effective field theory approach for Calabi-Yau compactifications was used more recently~\cite{Becker:2015wga}.  It is also the method used earlier for $G_2$ compactifications by K.Becker, D.Robbins and E.Witten~\cite{Becker:2015cca}.  The $\mathcal{N}=1$ sigma model approach for $G_2$ manifolds is work in progress~\cite{InProgress}.
\end{enumerate}
We shall see that the mathematical techniques used on the worldsheet and in space-time are very similar.  We begin with the worldsheet description of $\mathcal{N}=2$ models.

\section{Worldsheet approach: Nemeschansky and Sen}

Consider a $(2,2)$ nonlinear $\s$-model in superspace for which the target manifold is complex and K\"ahler,
\be
\int d^2\xi~ d^2\theta ~d^2\bar{\theta} K(\Phi,\widetilde{\Phi}).
\ee
Here $\Phi$ and $\widetilde{\Phi}$ are $\mathcal{N}=2$ superfields and $K$ is the K\"ahler potential of the target manifold.
To construct a theory of strings on a curved background it is necessary that worldsheet conformal invariance be preserved.  This implies that the worldsheet energy-momentum tensor is traceless
\be
T_i^{\hph{a}i}=0,
\ee
where $i,j=1,2$ describe worldsheet indices.
Equivalently, this condition can be formulated in terms of the target space metric $\beta$-function, see~\cite{Nemeschansky:1986yx} and references therein
\be
\beta^G_{ab}=R_{ab}+\alpha'\p_a\p_b\Delta \beta^K.
\ee
  Here $a,b$ denote again the six-dimensional real indices of the target manifold and the first term on the right hand side corresponds to the one-loop contribution, while the second term is the contribution from higher loops, which come in at order $\al'$ and higher. The form of the latter term follows from
    $\mathcal{N}=2$ supersymmetry, as the metric in this case can be expressed in terms of a single function, the K\"ahler potential $K$.

  Using complex coordinates $m,n,\cdots,\bar{m},\bar{n}=1,2,3$, the one-loop contribution is
\be
\beta^G_{mn}=R_{mn}=0,\qquad\beta^G_{\bar{m}\bar{n}}=R_{\bar{m}\bar{n}}=0,\non
\ee
\be
\beta^G_{m\bar{n}}=R_{m\bar{n}}=c\p_m\p_{\bar{n}}\Tr(\ln G).
\ee
Here $G$ is a $3\times 3$ complex matrix describing the target space metric and $c$ is a constant that is not relevant for our purpose.
The first two of the above equations vanish for K\"ahler manifolds.
The third equation says that the one-loop $\beta$-function vanishes for Ricci-flat manifolds. Thus $M_6$ is a Calabi-Yau manifold.

What happens once higher order corrections are included?
Define $\Delta\beta^K$ as the contribution to the $\beta$-function from loops $\ell\ge 2$ and
$\d K=K-\widetilde{K}$ as well as \qquad$\d G_{m\bar{n}}=G_{m\bar{n}}-\widetilde{G}_{m\bar{n}}=\p_m\p_{\bar{n}}\d K$.
Here the tilde denotes the Ricci-flat metric.

 The goal is to solve the equation of vanishing $\beta$-function to all orders in $\a'$
\be
c\p_m\p_{\bar{n}}\Tr(\ln G)+{\al'}\p_m\p_{\bar{n}}\Delta\beta^K(G)=0.
\ee
If the metric satisfies the condition
\be
c\Tr(\ln G)+\al'\Delta\beta^K(G)=c\Tr(\ln\widetilde{G}),
\ee
it certainly solves the former equation.
Equivalently this can be written as
\be
\label{eq:13}
\widetilde{G}^{m\bar{n}}\p_m\p_{\bar{n}}\d K=-\al'c^{-1}\Delta\beta^K(G)+\sum_{k=2}^\infty\frac{\lp -1\rp^k}{k}\Tr\lp\widetilde{G}^{-1}\d G\rp^k,
\ee
as can be easily verified.
Equation (\ref{eq:13}) can be solved iteratively for $\d K$. Since $\d K$ is of order $\al'$, the leading equation is
\be
\label{eq:14}
\widetilde{\square}\d K=-\al'c^{-1}\Delta\beta^K(\widetilde{G}).
\ee
$\Delta\beta^K(\widetilde{G})$ is a globally defined scalar (see \cite{Nemeschansky:1986yx} for a very detailed discussion on this non-trivial point, which holds beyond one-loop; at one-loop the corresponding $\Tr(\ln G)$ is not globally defined).
 Thus, a simple argument involving the Hodge decomposition theorem says one can separate it into a globally defined harmonic zero form $a_0$, that on a compact manifold is constant and a piece orthogonal to this zero mode, $b_0$, that is a globally defined zero form
\be
\Delta\beta^K(\widetilde{G})=a_0+\widetilde{\square}b_0.
\ee
Taking into account
\be
\widetilde{\square}\widetilde{K}=\widetilde{G}^{m\bar{n}}\p_m\p_{\bar{n}}\widetilde{K}=\widetilde{G}^{m\bar{n}}\widetilde{G}_{m\bar{n}}=3,
\ee
the solution to (\ref{eq:14}) is
\be
\d K=-\al'c^{-1}\lp\frac{a_0}{3}\widetilde{K}+b_0\rp.
\ee
The resulting metric
\be
\label{eq:NewMetric}
G_{m\bar{n}}=\widetilde{G}_{m\bar{n}}-\al'c^{-1}\lp\frac{a_0}{3}\widetilde{G}_{m\bar{n}}+\p_m\p_{\bar{n}}b_0\rp,
\ee
is again K\"ahler but not Ricci-flat. Note that the metric belongs to the same K\"ahler class as a Ricci flat metric (which is the leading order Ricci-flat metric rescaled by the constant factor $(1-\alpha'c^{-1}a_0/3)$) and that $c_1(M_6)=0$. The corrected manifold $M_6'$ is Calabi-Yau in the topological sense with a representative that is Ricci-flat in the same K\"ahler class.

Note that 
\be
\d G_{m\bar{n}}=-\al'c^{-1}\lp\frac{a_0}{3}\widetilde{G}_{m\bar{n}}+\p_m\p_{\bar{n}}b_0\rp,
\ee
is a globally defined tensor field, as $b_0$ is a globally defined scalar field. Thus the new metric
\be
G_{m\bar{n}}=\widetilde{G}_{m\bar{n}}+ \d G_{m\bar{n}},
\ee
is an admissible metric on the Calabi-Yau manifold.
This process can be performed iteratively by inserting
\be
\d G_{m\bar{n}}=\p_m\p_{\bar{n}}\d K,
\ee
into the right hand side of (\ref{eq:13}).  The resulting expression is a globally defined scalar field because we had just seen, that $\d G_{m\bar{n}}$ is a globally defined tensor. This is the only input used to arrive at the solution (2.13), so the new equations may be solved as before to find an all orders solution.  At each step we will get an equation like (\ref{eq:NewMetric}), showing that the new metric which will be in the same K\"ahler class as a Ricci-flat metric (which is a constant rescaling of the original metric).

\section{Space-time approach}

In this framework we can make contact with the beautiful mathematics of $G$ structures, so it puts the previous worldsheet approach into a slightly different perspective.
Of course, all features of the corrected target manifold $M'_6$ emerge in this approach as well
\begin{enumerate}
\item $M'_6$ is K\"ahler.
\item $c_1(M'_6)=0$.
\item $R_{ab}\ne 0$ with $a,b=1,\cdots,6$, so $M'_6$ is no longer Ricci-flat.
\end{enumerate}
As an additional bonus, we show that $M'_6$ has $\SU(3)$ structure rather than $\SU(3)$ holonomy.

\subsection{The $\SU(3)$ structure}

There is an excellent mathematical literature defining rigorously the concept of (the more general) $G$ structure (see e.g.~\cite{MR1787733}).  For our purpose we define an $\SU(3)$ structure as a collection of $\SU(3)$ invariant globally defined (real) forms $J$, $\Om_1$, $\Om_2$,
\be
\lp J,\Om_1,\Om_2\rp.
\ee
Here $J$ is a two-form and $\Om_1$, $\Om_2$ are three-forms.  In the Calabi-Yau case these forms are closed, but this is no longer (necessarily) true for $\SU(3)$ structure manifolds.  Failure of the structure group $SU(3)$ to be the holonomy group $SU(3)$  is characterized by the components of the intrinsic torsion
\begin{equation}
\begin{split}
& d J = - {3\o 2} {\rm Im} (W_1  \bar \Omega) + W_4 \wedge J + W_3, \\
& d \Omega = W_1 J^2 + W_2 \wedge J + \bar W_5 \wedge \Omega.
\end{split}
\end{equation}
see e.g.~\cite{Kaste:2003dh} or one of the many other references on $SU(3)$ structure manifolds. The explicit form of the torsion classes $W_i$ is not needed in the following though.

An $\SU(3)$ structure singles out a basis of (real and commuting) spinors on $M_6$,
\be
\lp\eta,i\G\eta,i\G_a\eta\rp,
\ee
and a dual basis
\be
\lp\eta^T,-i\eta^T\G,-i\eta^T\G_a\rp.
\ee
Here $\G_a$ are the 6D gamma matrices previously introduced, that can be chosen to be imaginary and anti-symmetric
and  $\G=\frac{i}{6}\e^{abcdef}\G_{abcdef}$ is the 6D chirality operator.

Being a basis on $M_6$ these spinors satisfy the completeness relation or Fierz identity
\be
\label{eq:Completeness}
1=\eta\eta^T+\G\eta\eta^T\G+\G^a\eta\eta^T\G_a,
\ee
as can be easily verified.

The $SU(3)$ structure on $M_6$ can be constructed as spinor bilinears
\bea
J_{ab} &=& -i\eta^T\G_{ab}\G\eta,\non\\
\Om_{1\,abc} &=& -i\eta^T\G_{abc}\eta,\\
\Om_{2\,abc} &=& -\eta^T\G_{abc}\G\eta.\non
\eea
These forms are not independent but satisfy some relations which follow from the Fierz identity (\ref{eq:Completeness}),
\be
\label{eq:FormConstraints}
J\w\Om_1=J\w\Om_2=0,\qquad J\w J\w J=\frac{3}{2}\Om_1\w\Om_2.
\ee
There are more identities for the dual forms
\begin{equation}
\begin{split}
\star \Omega_1& = - \Omega_2, \\
\star J & = J \wedge J, \\
\end{split}
\end{equation}
see \cite{Becker:2015wga} for a complete list of equivalences.

Further, $\Om_2$ can be expressed in terms of the triple $(J,\Omega_1,g_{ab})$ as can be verified using again the Fierz identity
\be
\label{eq:Om2Relation}
\Om_{2\,abc}=-J_{ad}\Om_{1\,bce}g^{de}.
\ee
Finally, the metric follows from the pair $(J,\Om_1)$
\be
g_{ab}=g_{ab}[J,\Om_1],
\ee
as has been shown in  \cite{Becker:2015wga}, where it was verified that the proposed metric satisfies all the duality-, contraction-, and normalization-identities.
Such a relation between the metric and the $SU(3)$ structure appears for the $G_2$ case in e.g. the review ~\cite{Grigorian:2009ge}. The identity for the $SU(3)$ case is new, as far as we know.

In conclusion, we can solve the constraints (\ref{eq:FormConstraints}) along with the SUSY constraints (which follow next) for the pair $(J,\Om_1)$, while $\Om_2$ and the metric $g_{ab}$ follow by the previous relations.

\subsection{Supersymmetry constraints}

The covariant derivative of a normalized spinor $\eta$ can be expressed (very generally) in terms of the spinor basis
\be
\nabla_a\eta=iA_a\G\eta+iB_a^{\hph{a}b}\G_b\eta.
\ee
Here $A_a$ and $B_a^{\hph{a}b}$ are matrices encoding the $\al'$-corrections to the gravitino SUSY transformation.  We do not need the explicit form of these corrections except for some of their properties.  To leading order in $\al'$ we have $A_a=B_a^{\hph{a}b}=0$ (as previously discussed), while to order ${\al'}^3$ the corrections are explicitly known (see refs. in~\cite{Becker:2015wga}).

Taking covariant derivatives of the $\SU(3)$ structure,
\be
\nabla_aJ_{bc}=-2i\eta^T\G_{bc}\G\nabla_a\eta=2B_a^{\hph{a}b}\Om_{2\,bcd},
\ee
we observe that $B_a^{\hph{a}b}=0$ for the manifold $M_6$ to remain K\"ahler. Non-K\"ahler manifolds lead to time dependent solutions once the space-time fields are taken into account, as pointed out by Gross and Witten~\cite{Gross:1986iv}.  The covariant derivative of $\Om_1$ is
\be
\nabla_a\Om_{1\,bcd}=-2i\eta^T\G_{bcd}\nabla_a\eta=-2A_a\Om_{2\,bcd}.
\ee

The antisymmetrized equations take the form
\be
\label{eq:StructureEquations}
dJ'=0\qquad\mathrm{\&}\qquad d\Om_1'=-2A\w\Om_2'.
\ee
If the right hand side of the second equation is non-vanishing, we observe that $\Om_1'$ is no longer closed. The manifold then fails to have $SU(3)$ holonomy, but rather has $SU(3)$ structure.
The goal is to solve the previous equations along with the constraints
\be
\label{eq:GeneralConstraints}
J'\w\Om_1'=0\qquad\mathrm{\&}\qquad J'\w J'\w J'=\frac{3}{2}\Om_1'\w\Om_2'.
\ee
Here the prime denotes the $\al'$-corrected forms
\be
\label{eq:CorrectedForms}
J'=J+\d J\qquad\mathrm{\&}\qquad\Om_1'=\Om_1+\d\Om_1,
\ee
and $\Om_2'$ is constructed from $J'$ and $\Om_1'$ as in (\ref{eq:Om2Relation}).

 These equations can be solved iteratively in $\al'$.
A convenient ansatz for the first order corrected  $\SU(3)$ structure, that uses input from group theory (see~\cite{Becker:2015wga} for details) is
\be
J'=J+da\qquad\mathrm{\&}\qquad\Om_1'=\Om_1-2\la\Om_2+2x\Om_1+db.
\ee
Here $\la$ and $x$ describe one forms encoding the $\al'$-corrections, that are assumed to be known
\be
A_a=\nabla_a\l+{J_a}^b\nabla_bx.
\ee
Further, $a$ and $b$ denote a one form and a two form respectively.
The leading order result to the equations (\ref{eq:StructureEquations}) satisfying the constraints (\ref{eq:GeneralConstraints}) is
\be
a_a=J_a^{\hph{a}b}\nabla_b\rho\qquad\mathrm{\&}\qquad b_{ab}=0,
\ee
Here $\rho$ satisfies the Poisson equation
\be
\nabla^2\rho+4x=0.
\ee
This can always be solved, because $x$ is determined only up to a constant $c$
\begin{equation}
0 = \int _{M_6'} \nabla^2 \rho = -4 \int_{M_6'} x + c.
\end{equation}

The equations (\ref{eq:StructureEquations}) and (\ref{eq:GeneralConstraints}) can then be evaluated to the next order and beyond to find an all orders solution.

\section{Conclusion}

To summarize: the space-time approach leads to the same conclusions about $M_6'$ as the worldsheet approach, just from a different perspective.
\begin{enumerate}
\item $M_6'$ remains K\"ahler but is no longer Ricci-flat.
\item The corrected manifold $M_6'$ has a vanishing first Chern class, $c_1(M_6')=0$,
\end{enumerate}
so $M_6'$ is Calabi-Yau in the topological sense, but not in the metric sense.
The space-time approach showed, that $M_6'$ has no longer $\SU(3)$ holonomy but rather $\SU(3)$structure,
because the holomorphic three-form is no longer closed.

\section*{Acknowledgement}\addcontentsline{toc}{section}{Acknowledgement}

We would like to thank A. Sen for helping us clarify some of our confusions about the worldsheet appoach. This work was supported by the National Science Foundation grant PHY-1214344.



\providecommand{\href}[2]{#2}\begingroup\raggedright\endgroup

\end{document}